\documentclass{article}
\usepackage{microtype}
\usepackage{graphicx}
\usepackage{subfigure}
\usepackage{booktabs} 
\usepackage{enumitem}
\usepackage{hyperref}
\usepackage[T1]{fontenc}    
\usepackage{url}          
\usepackage{amsfonts} 
\usepackage{amsmath}

\usepackage[accepted]{icml2021}
\renewcommand{\algorithmiccomment}[1]{#1} 

\icmltitlerunning{HEMET: A Homomorphic-Encryption-Friendly Privacy-Preserving Mobile Neural Network Architecture}
\begin{document}
	
	\twocolumn[
	\icmltitle{HEMET: A Homomorphic-Encryption-Friendly Privacy-Preserving Mobile Neural Network Architecture}
	
	\begin{icmlauthorlist}
		\icmlauthor{Qian Lou}{in}
		\icmlauthor{Lei Jiang}{in}

		\icmlkeywords{Homomorphic Encryption, Mobile Network Architecture}
	\end{icmlauthorlist}
	
	\icmlaffiliation{in}{Indiana University Bloomington}
	\icmlcorrespondingauthor{Lei Jiang}{jiang60@iu.edu}
	
	\vskip 0.3in
	]
	
\printAffiliationsAndNotice{}	

\begin{abstract}
Recently Homomorphic Encryption (HE) is used to implement Privacy-Preserving Neural Networks (PPNNs) that perform inferences directly on encrypted data without decryption. Prior PPNNs adopt mobile network architectures such as SqueezeNet for smaller computing overhead, but we find na\"ively using mobile network architectures for a PPNN does not necessarily achieve shorter inference latency. Despite having less parameters, a mobile network architecture typically introduces more layers and increases the HE multiplicative depth of a PPNN, thereby prolonging its inference latency. In this paper, we propose a \textbf{HE}-friendly privacy-preserving \textbf{M}obile neural n\textbf{ET}work architecture, \textbf{HEMET}. Experimental results show that, compared to state-of-the-art (SOTA) PPNNs, HEMET reduces the inference latency by $59.3\%\sim 61.2\%$, and improves the inference accuracy by $0.4 \% \sim 0.5\%$.
\end{abstract}

\section{Introduction}
\label{s:intro}
	
Clients feel reluctant to upload their sensitive data, e.g., health or financial records, to untrusted servers in the cloud. To protect clients' privacy, privacy-preserving neural networks (PPNNs)~\cite{GAZELLE:USENIX18,mishra2020delphi,Brutzkus:ICML2019,Gilad-Bachrach:ICML2016,dathathri:2019PLDI} are built to perform inferences directly on encrypted data. An interactive PPNN~\cite{GAZELLE:USENIX18,mishra2020delphi} uses homomorphic encryption (HE) for linear layers, and adopts secure multi-party computation (MPC) to process activation layers. However, huge volumes of data between the client and the server have to be exchanged during an inference of an interactive PPNN. For instance, DELPHI~\cite{mishra2020delphi} has to transmit 2GB data for only a ResNet-32 inference on a single encrypted CIFAR-10 image. On the contrary, non-interactive PPNNs~\cite{Brutzkus:ICML2019,dathathri:2019PLDI,Dathathri:PLDI20:EVA} approximate their activations by a degree-2 polynomial, and compute an entire inference via HE. They do not require high network bandwidth, but still can obtain competitive inference accuracy~\cite{Dathathri:PLDI20:EVA}. Thereafter (except Section~\ref{s:ppnn_o}), when we mention a PPNN, we indicate a non-interactive PPNN.

Unfortunately, PPNN inferences are time-consuming. An inference of a typical PPNN~\cite{dathathri:2019PLDI} requires $>2$ seconds on an encrypted MNIST image, and consumes $>70$ seconds on an encrypted CIFAR-10 image. There is a $\times 10^6$ latency gap between a PPNN inference and a unencrypted inference. To mitigate the gap, recent work~\cite{Dathathri:PLDI20:EVA,dathathri:2019PLDI} adopts mobile neural network architectures such as SqueezeNet~\cite{iandola2016squeezenet} and InceptionNet~\cite{szegedy2016inception} to implement PPNNs. However, we find na\"ively adopting mobile neural network architectures for PPNNs does not necessarily achieves shorter inference latency. Mobile neural network models~\cite{szegedy2016inception,iandola2016squeezenet} reduce the total number of parameters but still maintain competitive inference accuracy by adding more linear layers. In spite of less parameters, if a PPNN adopts a mobile neural network architecture, its deeper architecture with more layers greatly increases the HE \textit{multiplicative depth}, thereby decelerating each HE operations of the PPNN, where the multiplicative depth means the number of HE multiplications on the critical path.


In this paper, we propose a Homomorphic-Encryption-friendly privacy-preserving Mobile neural network architecture, HEMET, to achieve shorter inference latency and higher inference accuracy. Our contributions can be summarized as the following.
\begin{itemize}[noitemsep,topsep=0pt,parsep=0pt,partopsep=0pt,leftmargin=*]

\item We first identify that na\"ively applying a mobile neural network architecture on a PPNN may even prolong its inference latency. Although the mobile network architecture reduces HE operations of the PPNN, but it greatly increases the multiplicative depth and decelerates each HE operation of the PPNN.

\item We propose a simple, greedy HE-friendly mobile network architecture search algorithm to evaluate whether a block should adopt a regular convolutional layer or a mobile module to minimize the inference latency of the entire network. The search algorithm is performed layer by layer, and can reduce HE operations without increasing the multiplicative depth of a PPNN.

\item We also present Coefficient Merging to further reduce the multiplicative depth of a PPNN by merging the mask, approximated activation coefficients, and batch normalization coefficients of each layer.
		
\item We evaluated and compared HEMET against SOTA PPNN architectures. Our experimental results show HEMET reduces the inference latency by $59.3\%\sim 61.2\%$, but improves the inference accuracy by $0.4\% \sim 0.5\%$ over various prior PPNNs.

\end{itemize}

\begin{table}[t!]
\small
\centering
\setlength{\tabcolsep}{4pt}
		\begin{tabular}{|c||c|} \hline
			HE Operations               & CKKS-RNS  $Q=\prod_{i=1}^{r}{Q_i}$ \\\hline\hline
			Addition, Subtraction       & $\mathcal{O}(N\cdot r)$         \\\hline
			Scalar Multiplication		& $\mathcal{O}(N\cdot r)$         \\\hline
			Plaintext Multiplication    & $\mathcal{O}(N\cdot r)$         \\\hline
			Ciphertext Multiplication	& $\mathcal{O}(N\cdot log(N)\cdot r^2)$         \\\hline
			Ciphertext Rotation	        & $\mathcal{O}(N\cdot log(N)\cdot r^2)$         \\\hline	
		\end{tabular}
		\caption{The complexity of HE operations on a $r$-level ciphertext ($N$ is the polynomial degree of the ciphertext).}
		\label{t:moti2_merge}
\vspace{-0.2in}
\end{table}

\section{Background}
\label{s:back}

\subsection{Homomorphic Encryption}
HE allows arbitrary computations to occur on encrypted data (ciphertexts) without decryption. Given a public key $pk$, a secret key $sk$, an encryption function $\epsilon ()$, and a decryption function $\sigma ()$, a HE operation $\otimes$ can be defined if there is another operation $\times$ such that $\sigma(\epsilon(x_1, pk)\otimes\epsilon(x_2, pk), sk) = \sigma(\epsilon(x_1\times x_2, pk), sk)$, where $x_1$ and $x_2$ are plaintexts, each of which encodes a vector consisting of multiple integer or fixed-point numbers~\cite{dathathri:2019PLDI, Dathathri:PLDI20:EVA}. Each HE operation introduces a certain amount of noise into the ciphertext. When the accumulated noise grows beyond a noise budget, a HE decryption failure happens. Though a \textit{bootstrapping} operation~\cite{Gentry:HE2009} reduces the accumulated noise in a ciphertext, it is computationally-expensive. Prior PPNNs use \textit{leveled} HE having a fixed noise budget to compute only a limited number of HE operations without bootstrapping.

\subsection{Privacy-Preserving Neural Networks}
\label{s:ppnn_o}

Recent PPNNs~\cite{GAZELLE:USENIX18,mishra2020delphi,Brutzkus:ICML2019,Gilad-Bachrach:ICML2016,dathathri:2019PLDI,Dathathri:PLDI20:EVA} adopt HE to implement their linear layers. Unfortunately, HE cannot support non-linear activation layers. Interactive PPNNs~\cite{GAZELLE:USENIX18,mishra2020delphi} take advantage of multi-party computation to compute activation by interactions between the client and the server. In contrast, non-interactive PPNNs~\cite{Brutzkus:ICML2019,Gilad-Bachrach:ICML2016,dathathri:2019PLDI,Dathathri:PLDI20:EVA} approximate activations by a degree-2 polynomial, e.g., square function, so that the entire PPNN inference happens on only the server. Non-interactive PPNNs are more friendly to the clients who have no powerful machines and high-bandwidth network connections. The latest interactive PPNN, Delphi~\cite{mishra2020delphi}, has to exchange 2GB data between the server and the client for only a ResNet-32 inference on an encrypted CIFAR-10 image. In this paper, we focus on only non-interactive PPNNs.

\subsection{Threat Model}
The threat model of HEMET is similar to that of prior PPNNs~\cite{Brutzkus:ICML2019,Gilad-Bachrach:ICML2016,dathathri:2019PLDI,Dathathri:PLDI20:EVA}. Though an encryption scheme can be used to encrypt data sent to cloud, untrusted servers can make data leakage happen. HE enables a server to perform private inferences over encrypted data. A client sends encrypted data to a server performing encrypted inferences without decrypting the encrypted data or accessing the client's secret key. Only the client can decrypt the inference results using the secret key.

\begin{figure*}[t!]
	\centering
  \includegraphics[width=6.3in]{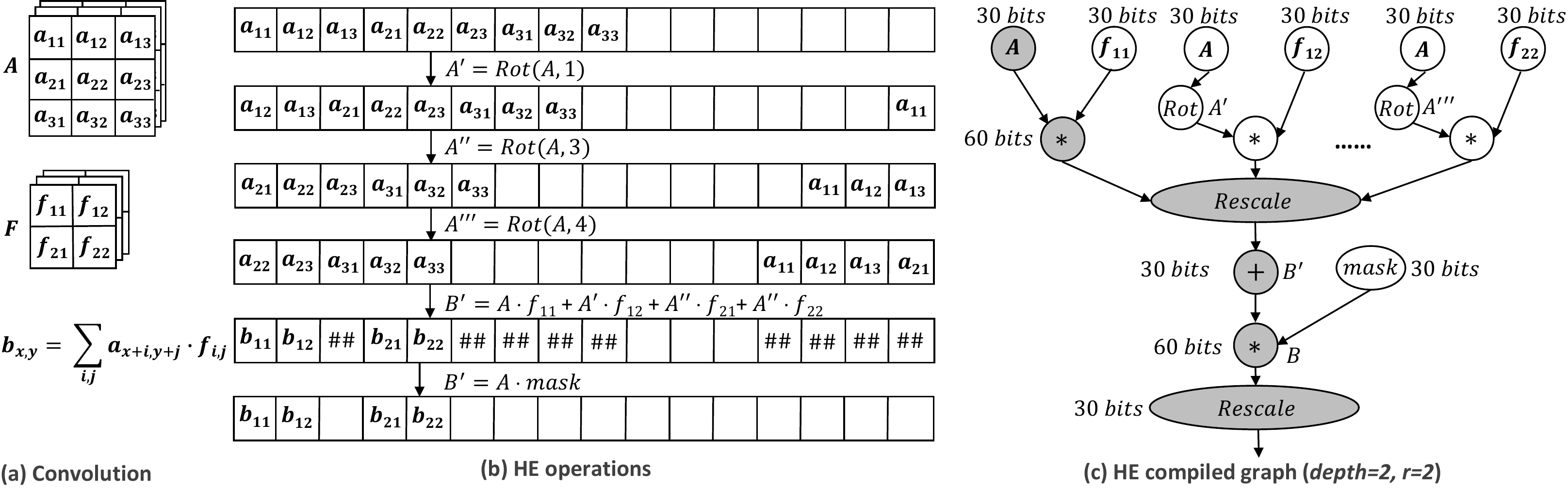}
	\vspace{-0.1in}
	\caption{A CKKS-based PPNN.}
	\label{f:ckks_ppnns}
	\vspace{-0.15in}
\end{figure*}

\begin{figure*}[t!]
		\centering
		\includegraphics[width=6.6in]{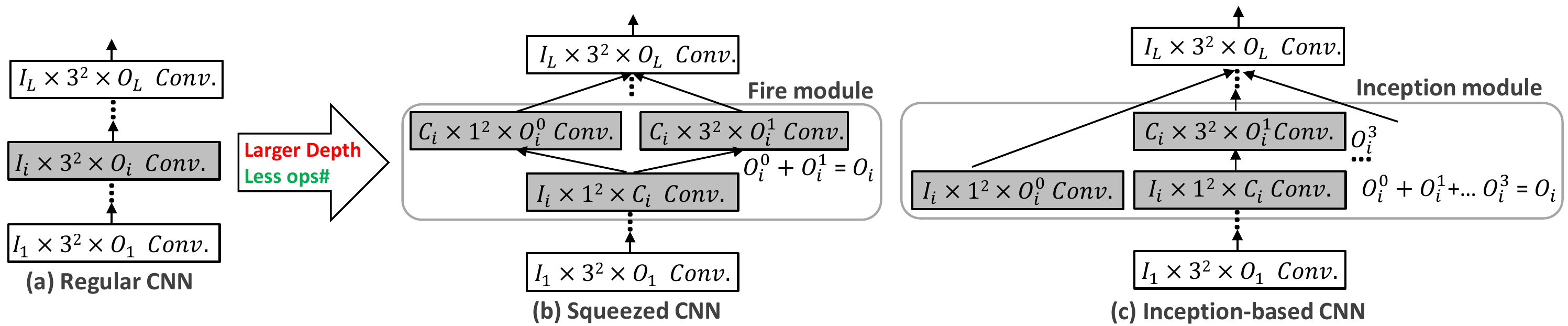}
		\vspace{-0.1in}
		\caption{Various mobile network architectures.}
		\label{f:prior_networks}
		\vspace{-0.2in}
\end{figure*}

\subsection{A RNS-CKKS-based PPNN}

\textbf{RNS-CKKS Scheme}. Among all HE schemes, RNS-CKKS is the \textbf{only} scheme that supports fixed-point arithmetic operations. Recent PPNNs~\cite{dathathri:2019PLDI,Dathathri:PLDI20:EVA} use RNS-CKKS to achieve shorter inference latency and higher inference accuracy than the PPNN~\cite{Gilad-Bachrach:ICML2016} implemented by other HE schemes. Via SIMD batching, a RNS-CKKS-based PPNN encrypts $\frac{N}{2}$ fixed-point numbers in $\frac{N}{2}$ slots of a single ciphertext, where $N$ is the polynomial degree of the ciphertext. One HE operation on the ciphertext simultaneously performs the same operation on each slot of the ciphertext. A ciphertext is a polynomial of degree $N$ with each coefficient represented modulo $Q$, where $Q$ is a product of $r$ primes $Q=\prod_{j=1}^{r}{Q_j}$. To represent a fixed-point number, RNS-CKKS uses an integer $I$ and a scaling factor $S$. For instance, 3.14 can be denoted by $I=314$ and $S=100$. $S$ grows exponentially with HE multiplication. To keep $S$ within check, a \textit{rescaling operation} is required to convert $I\times100$ at scale $S$ to $I$ at scale $S\times100$. Totally, RNS-CKKS permits $r$ rescaling operations. A rescaling operation converts a $r$-level ciphertext $C_r$ (which has noise $e$, modulus $Q$, and plaintext $m$) to a $(r-1)$-level ciphertext $C_{r-1}$ (which has noise $\frac{e}{Q_r}$, modulus $\frac{Q}{Q_r}$, and plaintext $\frac{m}{Q_r}$). $(C_{r-1}, \frac{m}{Q_r}, \frac{e}{Q_r}, \frac{Q}{Q_r})=Rescale(C_r, m, e, Q)$. The computational complexity of various RNS-CKKS operations on $r$-level ciphertexts is shown in Table~\ref{t:moti2_merge}. A rescaling operation makes the following operations faster, since the latency of a HE operation on a $(r-1)$-level ciphertext $C_{r-1}$ is shorter than that of the same operation on the $r$-level ciphertext $C_{r}$. Moreover, a rescaling operation also reduces noise in the ciphertext.

\textbf{RNS-CKKS-based Convolutions}. RNS-CKKS-based convolutions between an encrypted input tensor ($A$) and multiple plaintext weight filters ($F$s) are shown in Figure~\ref{f:ckks_ppnns}(a). The convolution result can be denoted as $b_{x,y} = \sum_{i,j}{a_{x+i,y+j}}\cdot f_{i,j}$. We assume the input $A$ has $C$ channels, a height of $H$, and a width of $W$. As Figure~\ref{f:ckks_ppnns}(b) shows, the input tensor can be encrypted into $C$ ciphertexts, each of which packs $H\times W$ input elements~\cite{dathathri:2019PLDI}. This is the $HW$ batching scheme. Or to fully utilize the slots in a ciphertext, $C\times H\times W$ input elements can be packed into $\lceil\frac{2\times C\times H\times W}{N}\rceil$ ciphertexts~\cite{dathathri:2019PLDI}, where $\lceil \rceil$ represents a rounding up operation. This is the $CHW$ batching scheme. To compute $a_{x+i,y+j}\cdot f_{i,j}$, element-wise HE multiplications happen between the input tensor and each weight filter. A HE rotation operation $rot(A, s)$ is required to align the corresponding data to support the HE accumulation, where $s$ represents the stride distance. At last, a plaintext mask is used to remove the irrelevant slots $\#\#$.


\textbf{The Critical Path of a DDG}. To precisely control noise, a recent RNS-CKKS compiler~\cite{Dathathri:PLDI20:EVA} converts an entire convolution into a data dependency graph (DDG) shown in Figure~\ref{f:ckks_ppnns}(c), and then inserts rescaling operations. We assume both inputs and weight filters use 30-bit fixed-point numbers. The result of a HE addition is still 30-bit, but each HE multiplication yields a 60-bit result. A rescaling operation rescales a multiplication result back to a 30-bit fixed-point number. The critical path (gray nodes) of the DDG has the multiplicative depth of 2. And the convolution example requires two rescaling operations. The multiplicative depth of a PPNN is the total number of HE multiplications along the critical path of its DDG.

\subsection{Mobile Neural Network Architecture}
	
In plaintext domain, mobile neural network architectures such as SqueezeNet~\cite{iandola2016squeezenet} and InceptionNet~\cite{szegedy2016inception} are built to reduce network parameters and inference overhead. The topologies of SqueezeNet and Inception are highlighted in Figure~\ref{f:prior_networks}.

\begin{itemize}[noitemsep,topsep=0pt,parsep=0pt,partopsep=0pt,leftmargin=*]

\item \textbf{SqueezeNet}. RNS-CKKS-based PPNNs~\cite{dathathri:2019PLDI,Dathathri:PLDI20:EVA} adopt the architecture of SqueezeNet~\cite{iandola2016squeezenet} for fast private inferences on CIFAR-10 images. SqueezeNet achieves similar accuracy to AlexNet using 50$\times$ fewer parameters. SqueezeNet replaces conventional convolution layers by fire modules shown in Figure~\ref{f:prior_networks}(b). Assume a conventional convolution layer has the dimension of $I_i\times 3 \times 3 \times O_i $, where $I_i$ is the input channel number, $3$ is the weight filter size, and $O_i$ are the output channel number in the $i$-th layer. On the contrary, a fire module consists of stacked convolution layers. $C_i$ ($<O_i$) is the output channel number of the first convolutional layer. $O_{i}^{0}+O_{i}^{1}=O_i$ is enforced to make the $i_{th}$ layer output have $O_i$ channels. If the $HW$ batching technique is used, to compute a regular convolution layer, we need to do $\mathcal{O}(I_i\times 3 \times 3 )$ rotations, and $\mathcal{O}(I_i\times 3 \times 3\times O_i)$ HE multiplications. In contrast, to compute a fire module of SqueezeNet, we need to perform $\mathcal{O}(I_i\times 1 +C_i\times (3 \times 3+1))$ rotations, and $\mathcal{O}(I_i \times 1 \times 1 \times C_i+ C_i \times 1 \times O_i^0+ C_i \times 3 \times 3\times O_i^1)$ HE multiplications. However, \textit{fire modules of SqueezeNet greatly increase the layer number of the mobile neural network architecture}.

\item \textbf{InceptionNet}. InceptionNet~\cite{szegedy2016inception} achieves VGG-like accuracy using $\sim$7$\times$ fewer parameters. As Figure~\ref{f:prior_networks}(c) shows, InceptionNet is built through substituting conventional convolution layers with inception modules. An inception module is composed of four types of filters, each of them has output channels $O_{i}^{0\sim 3}$. Although inception modules greatly reduce model parameters, \textit{InceptionNet has to almost double the layer number, compared to the VGG CNN}.
\end{itemize}

\subsection{Motivation}

\begin{figure}[t!]
\centering
\includegraphics[width=2.7in]{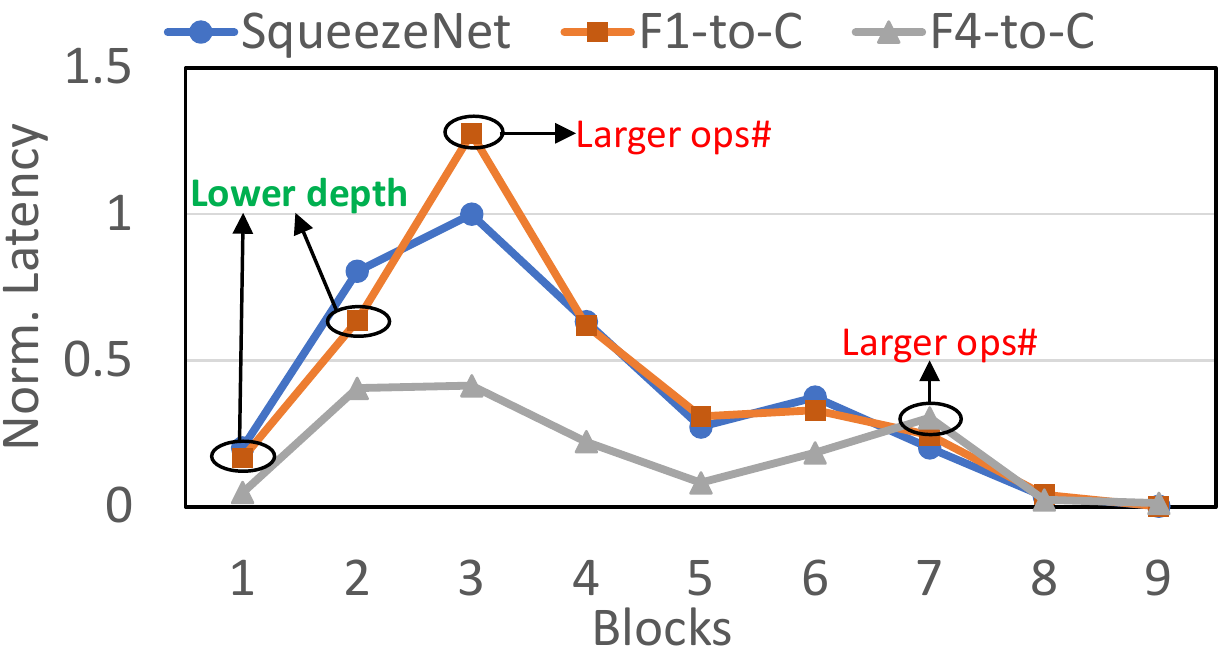}
\vspace{-0.1in}
\caption{Building a PPNN by fire modules of SqueezeNet.}
\label{f:moti_replacemnt}
\vspace{-0.2in}
\end{figure}

\textbf{A HE-friendly Mobile Network Architecture}. Though mobile neural networks such as SqueezeNet and InceptionNet reduce the number of HE operations of a PPNN, they significantly enlarge the multiplicative depth of the PPNN by adding more network layers. The enlarged multiplicative depth increases the computing overhead of each HE operation. As Figure~\ref{f:moti_replacemnt} shows, na\"ively adopting a mobile neural network architecture does not necessarily reduces the inference latency of a PPNN. A PPNN is built by sequentially connecting multiple \textit{block}s, each of which is either a fire module of SqueezeNet or a regular convolution layer. If the first fire module (F1) is replaced by a convolution layer having the same number of input and output channels, the inference latency (F1-to-C) of the PPNN increases by $\sim 32\%$. On the contrary, the inference latency (F4-to-C) of the PPNN reduces by $\sim 2 \times$, when the last fire module (F4) is replaced by a regular convolution layer. Therefore, we need a HE-friendly mobile network architecture to reduce the inference latency, and to maintain SOTA inference accuracy of a PPNN.

\textbf{Further Reducing the Multiplicative Depth}. As Figure~\ref{f:method2} shows, to build a PPNN, a given neural network is compiled into a DDG composed of additions, multiplications, and rotations. The computing overhead of each HE operation of the PPNN is decided by the PPNN multiplicative depth, which is roughly equal to the number of HE multiplications along the the critical (longest) path of the DDG. A multiplicative depth is defined by both the polynomial degree $N$ and the number of rescaling operations $r$, as shown in Table~\ref{t:moti2_merge}. The polynomial degree $N$ is determined for a fixed security level. A smaller $N$ may hurt the security level, so we aim to reduce the multiplicative depth and to accelerate HE operation by reducing the number of rescaling operations $r$. The larger $r$ is, the slower each HE operation is. To reduce $r$ and speedup each HE operation, the critical path of the DDG has to be shortened.

\subsection{Related Work}

\textbf{HE-friendly PPNNs}. Recent work~\cite{Bian:ECAI2020,Lou:NIPS2020} creates reinforcement learning agents to search a competitive neural network architecture implemented by BFV. However, these agents consider only regular convolutional layers, but not low-cost mobile network modules. Moreover, BFV supports only integer arithmetic operations, and thus has difficulties in controlling the scale of HE multiplication results. Recent PPNNs adopt RNS-CKKS with rescaling to solve this issue.

\textbf{RNS-CKKS Rescaling}. A recent RNS-CKKS-based PPNN~\cite{Dathathri:PLDI20:EVA} proposes a waterline-based rescaling technique to perform rescaling operations as late as possible. In this way, the number of rescaling operations required along the critical path of the DDG can be minimized. The waterline-based rescaling technique statically inserts rescaling operations to positions along the critical path. However, no prior work tries to shorten the critical path to minimize the number of rescaling operations.

\section{HEMET}

\subsection{Search Algorithm for a HE-Friendly Mobile Network Architecture}
\label{s:architecture_search}

We propose a simple, greedy search algorithm to find a HE-friendly mobile network architecture by performing block-wise evaluation on whether this block should be a regular convolutional layer or a mobile module consisting of multiple convolutional layers. We measure the latency difference between using a regular convolutional layer and adopting a mobile module. Only when the inference latency is reduced by the mobile module, we actually integrate the mobile module into the PPNN.

\textbf{Model Parameters and Multiplicative Depth}. Using a mobile network module, i.e., a fire or inception module, in a block of the PPNN reduces the HE operation in that block. Less HE operations might reduce the inference latency. However, the multi-layer mobile module also increases the number of layers in the PPNN, thereby enlarging the multiplicative depth of the PPNN. The enlarged multiplicative depth decelerates each HE operation in all layers before the next layer of the current mobile module by increasing the number of rescaling operations $r$ of the PPNN. It is critical to make sure that the benefit brought by a mobile network module is not offset by the deceleration caused by the increased multiplicative depth.

\setlength{\textfloatsep}{10pt}
\begin{algorithm}[tb!]
		\caption{HE-Friendly Network Architecture Search}
		\label{alg:replace}
		\begin{algorithmic}[1]
			\STATE {\bfseries Input:} Network $O$, Mobile module (block) number $n$
			\STATE {\bfseries Output:} HE-friendly network $O'$
			\STATE Initialize $O' = O$
			\FOR{$i=n$ {\bfseries to} $1$}
			\STATE  $Cost\_O' = compile\_run(O')$\\
			\COMMENT {//Replace the $i$-th module with a single Conv.}
			\STATE  $Tmp\_O' = replace\_back(O', i)$
			\STATE  $Cost\_Tmp\_O' = compile\_run(Tmp\_O' )$
			\IF{$Cost\_Tmp\_O' < Cost\_O'$}
			\STATE $O' = Tmp\_O'$
			\ENDIF
			\ENDFOR
		\end{algorithmic}
\end{algorithm}

\textbf{Search Algorithm}. The search algorithm for a HE-Friendly mobile network architecture is shown in Algorithm~\ref{alg:replace}. We assume $O$ is a mobile network consisting of only mobile modules and fully-connected layers. We first initialize the current architecture $O'$ to $O$, and compute its inference latency as $Cost\_O'$. We start to replace a mobile module by a regular convolutional layer from the last mobile module of $O'$, because the last mobile module is more likely to be replaced. If the last mobile module is replaced, the rescaling level of all previous mobile modules reduces, thereby greatly accelerating each HE operation in each mobile module. A candidate network architecture $Tmp\_O'$ is generated by replacing the current mobile module of $O'$  with a regular convolutional layer. And then, we compute the inference latency of $Tmp\_O'$ as $Cost\_Tmp\_O'$. We update $O'$ to $Tmp\_O'$ if $Cost\_Tmp\_O' < Cost\_O'$. The process repeats until $n$ mobile modules (blocks) are evaluated.

\begin{figure}[t!]
		\centering
		\includegraphics[width=3.3in]{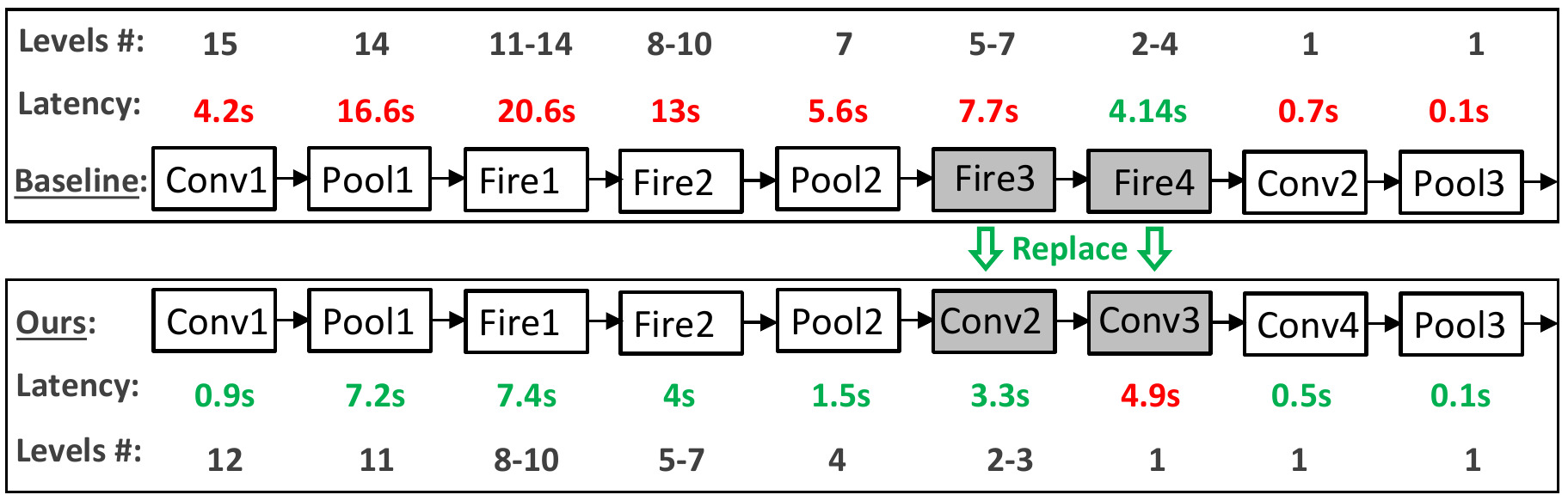}
		\vspace{-0.3in}
		\caption{HE-friendly network by layers replacement.}
		\label{f:method1}
\end{figure}

\textbf{A Case Study}. In Figure~\ref{f:method1}, we show a case study on how to find a HE-friendly mobile network architecture based on SqueezeNet by Algorithm~\ref{alg:replace}. SqueezeNet has four fire modules, and requires fifteen rescaling operations in the critical path. The HE operations in the first convolutional layer $Conv1$ happen on $15$-level of RNS-CKKS ciphertexts. Each HE operation manipulating the $15$-level ciphertexts is more computationally expensive than HE operations occurring on the lower level ciphertexts. SqueezeNet spends $72.7$ seconds in performing an inference on an encrypted CIFAR-10 image. By following Algorithm~\ref{alg:replace}, the third fire module with size of $I=64, C=32, O^0=128, O^1=128$ is replaced by a convolutional layer with input channel $I=64$, output channel $O=256$, and kernel size $k=3$. Similarly, the fourth fire module with size of $I=128, C=32, O^0=128, O^1=128$ is replaced by a convolutional layer with input channel $I=128$, output channel $O=256$, and kernel size $k=3$. The output mobile network requires only twelve rescaling operations, while SqueezeNet needs fifteen rescaling operations. Although the first convolutional layer $Conv1$ performs the same HE operations as SqueezeNet, it reduces the inference latency by $78\%$, since they happen on lower level ciphertexts.

\textbf{Algorithm Complexity and Security}. The asymptotic complexity of our network architecture search algorithm is $\mathcal{O}(n)$, which means that $n$ mobile modules need to be evaluated. As Figure~\ref{f:method1} shows, to build a HE-friendly network architecture based on SqueezeNet, totally four inferences have to be done on an encrypted CIFAR-10 image. The search overhead of Algorithm~\ref{alg:replace} is $\sum_{i=1}^{n}Cost\_O'_i$, where $O'_i$ is the $i_{th}$ network candidate. Our search algorithm is done offline before the PPNN is deployed on the server, so our algorithm does not expose any private information of input data. We maintain the same security level as~\cite{dathathri:2019PLDI,Dathathri:PLDI20:EVA}.

\subsection{Coefficient Merging}

\begin{figure}[t!]
		\centering
		\includegraphics[width=3.3in]{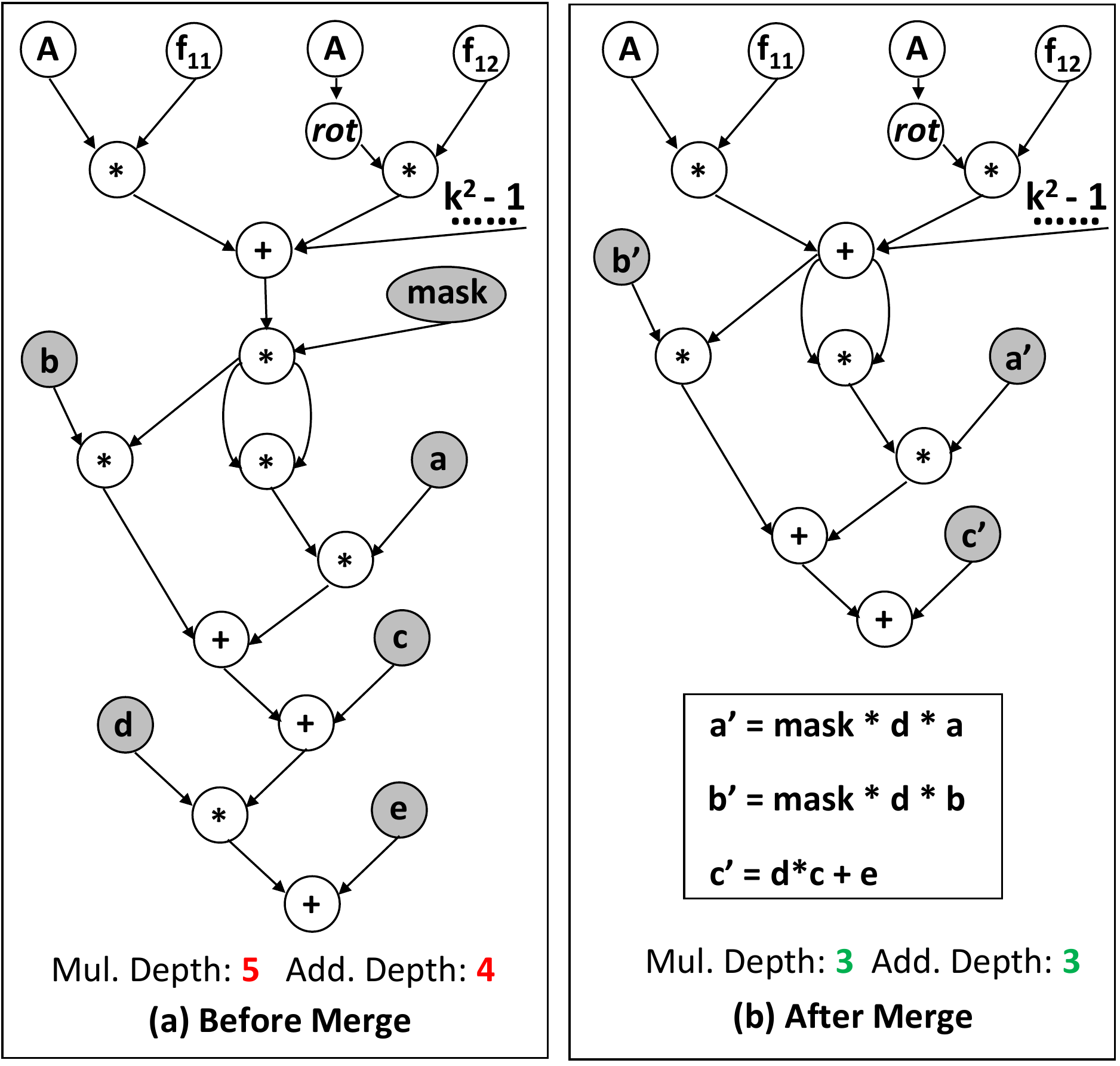}
		\vspace{-0.3in}
		\caption{Shortening the critical path by merging nodes.}
		\label{f:method2}
\end{figure}

To further reduce the inference latency of PPNNs, we propose \textit{Coefficient Merging} to reduce the length of the critical path of the DDG and the multiplicative depth of PPNNs. A PPNN having a larger multiplicative depth requires larger RNS-CKKS parameters to guarantee the 128-bit security level, thereby significantly increasing the computing overhead of each HE operation. Our search algorithm can find a HE-friendly mobile neural network architecture with less network layers. And then each layer of the resulting network can be compiled into a DDG shown in Figure~\ref{f:method2}, where each node is an operand or a HE operation. Coefficient Merging focuses on minimizing the multiplicative depth and reducing the number of rescaling operations $r$ by merging multiple nodes into one in the DDG.

\textbf{The DDG of a Convolutional Layer}. A convolution layer shown in Figure~\ref{f:ckks_ppnns}(a) can be described as Equation~\ref{e:convolution0}, where $A$ is the input, $f_{11}$ and $f_{22}$ are weight filers, and $X_{conv}$ is the convolution result. And $Y_{conv}$ is the result after removing wasted slots $\#\#$ from $X_{conv}$ by a mask $m$.   
\begin{equation}
\begin{split}
X_{conv} & = A f_{11} + rot(A,1) f_{12} \\
         & + rot(A,3) f_{21} + rot(A,3) f_{22}\\
Y_{conv} & = m X_{conv}
\label{e:convolution0}
\end{split}
\end{equation}	
A convolution is followed by an approximate degree-2 polynomial activation described as Equation~\ref{e:activation}, where coefficients $a$, $b$, and $c$ are learned in training. 
\begin{equation}
\label{e:activation}
Y_{act} = a Y_{conv}^2 + b Y_{conv} +c
\end{equation} 
Recent PPNNs adopt a batch normalization layer~\cite{Ioffe:ICML'15batchnorm} after each activation layer to improve inference accuracy. Batch normalization first calculates the mean $\mu$ and the variance $\sigma$ of the input of each layer, and then normalizes the input as $\overline{Y_{act}}$ using $\mu$, $\sigma$, and learned parameters $\gamma$ and $\beta$. 
\begin{equation}
\begin{split}
\label{e:batchnorm}
Y_{batch} & = \gamma \overline{Y_{act}} +\beta = \gamma  \frac{Y_{act}-\mu}{\sqrt{\sigma^2+\epsilon}}+\beta \\
Y_{batch} & = d Y_{act} +e
\end{split}
\end{equation}
In Equation~\ref{e:batchnorm}, we simplify the batch normalization layer using $d=\frac{\gamma}{\sqrt{\sigma^2+\epsilon}}$ and $e=\beta - \frac{\gamma \cdot \mu}{\sqrt{\sigma^2+\epsilon}}$. Figure~\ref{f:method2}(a) shows the DDG of a convolutional layer described by Equation~\ref{e:convolution0}$\sim$~\ref{e:batchnorm}. The convolution layer is followed by an approximate activation layer and a batch normalization layer.

\textbf{Coefficient Merging}. To reduce the critical path of the convolutional layer DDG (and the multiplicative depth of the PPNN), we can merge the mask $m$ in Equation~\ref{e:convolution0}, the coefficients $a$ and $b$ of the activation in Equation~\ref{e:activation}, and the coefficients $d$ and $e$ of the batch normalization in Equation~\ref{e:batchnorm}.
\begin{equation}
\label{e:batch_merge}
\begin{split}
	Y_{batch}  &= d(a m^2 X^2_{conv} + b m X_{conv} +c)+e\\
			       &= (d a m^2) X^2_{conv} + (d b m) X_{conv} +(d c +e) \\
			       &= a' X^2_{conv} + b' X_{conv} +c'
\end{split}
\end{equation} 
Equation~\ref{e:batch_merge} explains this merging process. Equation~\ref{e:batch_merge} merges $d\cdot a\cdot m$ into a single coefficient $a'$, so that two multiplications can be eliminated along the critical path. Equation~\ref{e:batch_merge} also merges $d \cdot b\cdot m$ and $d \cdot c +e$ into $b'$ and $c'$ respectively to further reduce HE multiplications in the DDG of a PPNN.

\textbf{Implementation}. Coefficient Merging can be implemented on a compiled DDG in three steps. First, we can generate a DDG for a specific network architecture. One example of a DDG for a convolution layer followed by an activation layer, and a batch normalization layer, is shown in Figure~\ref{f:method2}(a). Second, we parse the DDG to merge the mask, approximated activation coefficients, and batch normalization coefficients according to Equation~\ref{e:batch_merge}. If the convolutional layer is not followed by a batch normalization, we should remove the parts of $d$ and $e$ in Equation~\ref{e:batch_merge}. Batch normalization coefficients are learned from the training data of the server, but not from the input data of clients. Therefore, Coefficient Merging does not leak any information of clients. The last step of Coefficient Merging is to output an optimized DDG that has a smaller multiplicative depth and less HE operations than the original DDG. One example of the output of Coefficient Merging is shown in Figure~\ref{f:method2}(b). The multiplicative depth of the original DDG in Figure~\ref{f:method2}(a) is 5. Coefficient Merging significantly reduces the multiplicative depth of the DDG by $\sim40\%$. Furthermore, Coefficient Merging reduces the number $r$ of rescaling operations in each layer of a PPNN. For instance, Coefficient Merging can reduce $2\sim 3$ rescaling operations for SqueezeNet on CIFAR-10 dataset, leading to $20\%$ latency reduction. 

\textbf{Security}. Coefficient Merging does not leak any private information of clients. In our threat model where the server owns the neural networks model and the clients are the data owner, the goal is to prevent the untrusted server from accessing to the raw data of clients. In order to obtain higher inference accuracy and reduce inference latency, it is natural for the server with the original DDG, mask, approximated activation coefficients, and batch normalization coefficients, to perform Coefficient Merging before any PPNN service occurs. Coefficient Merging can be done offline, since the server does not require the inputs from clients.

	\begin{table*}[t!]
	\centering
	\footnotesize
		\setlength{\tabcolsep}{3pt}
		\begin{tabular}{|c||c|c|c|} \hline
			Network                 & Architecture                                                          & Accuracy (CIFAR-10) & Accuracy (CIFAR-100)\\\hline\hline
			AlexNet                 & C1-A1-P1-C2-A2-P2-C3-A3-C4-A4-C5-A5-P3-D1-A1-D2-A2-D3                 & 81.1\%        & 54.2\%  \\\hline
			SqueezeNet              & C1-P1-F1-F2-P2-F3-F4-C2-P3                                            & 81.5\%        & 65.3\%\\\hline	
			\textbf{Our\_F34}   & C1-P1-F1-F2-P2-C2-C3-C4-P3                                                & \textbf{81.9\%}   & 65.5 \%      \\\hline \hline
			InceptionNet            & C1-B1-I1-I2-P1-I3-I4-I5-I6-I7-I8-I9-P2-D1                             & 83.2\%         &69.1\%\\\hline	
			\textbf{Our\_I789} & C1-B1-I1-I2-P1-I3-I4-I5-I6-C3-C4-C5-P2-D1                                  & \textbf{83.7\%}      & 69.6\%      \\\hline
		\end{tabular}
		\caption{The network architecture of PPNNs. C$x$, A$x$, P$x$, D$x$, B$x$, F$x$, and I$x$ denote $x_{th}$ convolution, activation, pooling, dense, batch normalized activation, fire module, and inception module layers, respectively. More detailed parameters of each layer can be found in Appendix.}
		\label{t:Networks}
	\end{table*}

	\begin{table*}[t!]
	\centering
	\small
	\begin{tabular}{|l||c|c|c|c|c|c|c|} \hline
			Scheme                & Layer\# & rescale\# &N     &Q            & Latency                 & Accuracy (CIFAR-10)  & Accuracy (CIFAR-100)           \\\hline\hline
			EVA-AlexNet           & 11      & 14        &32768 & 820         & 293.5 seconds           & 81.1\%             & 54.2\%                      \\\hline
			EVA-SqueezeNet        & 13      & 17        &65536 & 1020        & 72.7 seconds            & 81.5\%             & 65.3\%                       \\\hline
			Ours-F4               & 12      & 15        &32768 & 880         & 43.8 seconds            & 81.7\%             & 65.3\%                       \\\hline
			\textbf{Ours-F34}     & 11      & 14        &32768 & 820         & 37.5 seconds            & 81.9\%             & 65.5\%                       \\\hline
			Ours-F234             & 10      & 13        &32768 & 760         & 50.2 seconds            & 81.9\%  			  & 65.6\%                        \\\hline
		\textbf{Ours-F34-Merge}   & 11      & 12        &32768 & 720         & \textbf{29.6 seconds}   & 81.9\%  			  &	65.5\%						\\\hline	
	\end{tabular}
	\caption{The inference latency and accuracy of SqueezeNet on CIFAR-10 and CIFAR-100. Ours-F$xyz$ indicates we replace the $x_{th}$, $y_{th}$, and $z_{th}$ fire module by a regular convolution layer. Ours-F$34$-Merge represents coefficient merging is applied on Ours-F$34$.}
	\label{t:SqueezeNet}
	\end{table*}

	\section{Experimental Methodology}

\subsection{Datasets and Networks}

\textbf{Datasets}. We adopt the datasets of CIFAR-10 and CIFAR-100 to evaluate our proposed techniques, because they are the most complex datasets prior PPNNs can be evaluated on~\cite{dathathri:2019PLDI,Dathathri:PLDI20:EVA}. The CIFAR-10 dataset includes 60K color images in 10 classes with the size of 32 $\times$ 32, where each class consists of 6K images. 50K images are used for training, while 10K images are used for testing in CIFAR-10. The CIFAR-100 dataset is similar to CIFAR-10, except it has 100 classes, each of which has 600 images. 

\textbf{Networks}. Table~\ref{t:Networks} shows the comparison between our baseline networks and HE-friendly networks. We first adopt AlexNet as our regular CNN baseline, and SqueezeNet as our mobile neural network baseline. We perform our search algorithm on SqueezeNet to find a faster network architecture Ours\_F34 with almost the same inference accuracy. Moreover, we also explore the design space of deeper mobile PPNN to achieve higher inference accuracy by InceptionNet. We find a new HE-friendly network architecture Ours\_I789 by searching on InceptionNet via Algorithm~\ref{alg:replace}.

\subsection{Experimental Setup}

We ran all PPNN inferences on a server-level hardware platform, which is equipped with an Intel Xeon Gold 5120 2.2GHz CPU with 56 cores and 256GB DRAM memory. Neural networks are trained by TensorFlow. For each neural network in Table~\ref{t:Networks}, we adopt the EVA compiler~\cite{Dathathri:PLDI20:EVA} to convert the neural network to a RNS-CKKS-based PPNN DDG. The EVA compiler is built upon the Microsoft SEAL library~\cite{sealcrypto}. By following~\cite{Dathathri:PLDI20:EVA}, we set the initial scale of encrypted input message to 25-bit, and the scale of weight filters and masks to 15-bit. The coefficients of approximated activation layers and batch normalization layers are set to 10-bit. All PPNNs in our experiments can achieve the 128-bit security level.

\section{Results and Analysis}

We report inference latency and accuracy of various PPNN architectures, layer numbers, rescaling operation numbers, and RNS-CKKS parameters (i.e., polynomial degree $N$ and modulus size $Q$) in Table~\ref{t:SqueezeNet} and~\ref{t:InceptionNet}. The datasets of CIFAR-10 and CIFAR-100 use the same PPNN architecture but with different weight filter numbers. EVA~\cite{dathathri:2019PLDI} is the SOTA FHE compiler that can convert a plaintext network model to a RNS-CKKS-based PPNN model. We compare HEMET against EVA~\cite{dathathri:2019PLDI} on the datasets of CIFAR-10 and CIFAR-100 by various network architectures. Table~\ref{t:SqueezeNet} and~\ref{t:InceptionNet} show the comparison between EVA and HEMET on SqueezeNet and InceptionNet, respectively.

	\begin{table*}[hbt!]
	\centering
	\small
	\begin{tabular}{|l||c|c|c|c|c|c|c|c|} \hline
		Scheme                & Layer\# & rescale\# &N     & Q bits       & Latency & Accuracy (CIFAR-10) & Accuracy (CIFAR-100)\\\hline\hline
		EVA-InceptionNet          & 34      & 39        &131072& 2340         & 213.2 seconds   & 83.2\%              & 69.1\% \\\hline
		Ours-I9               & 31      & 36        &131072 & 2160      &  173.5 seconds & 83.7\%    & 69.1\%\\\hline
		\textbf{Ours-I789}             & 28      & 33        &131072 & 1980    &  \textbf{132.8 seconds} & 83.7\% & 69.6\% \\\hline
		Ours-I6789            & 26      & 30        &131072 & 1980      &  193.6 seconds & 83.7\%  & 69.8\%\\\hline
		\textbf{Ours-I789-Merge}       & 19      & 29        &65536 & 1740       &  \textbf{83.2 seconds} & 83.7\% & 69.6\%\\\hline
	\end{tabular}
	\caption{The inference latency and accuracy of InceptionNet on CIFAR-10 and CIFAR-100. Ours-I$xyz$ indicates we replace the $x_{th}$, $y_{th}$, and $z_{th}$ inception module by a regular convolution layer. Ours-I$789$-Merge represents coefficient merging is applied on Ours-I$789$.}
	\label{t:InceptionNet}
	\end{table*}

\subsection{SqueezeNet}

As Table~\ref{t:SqueezeNet} shows, the PPNN of AlexNet generated by EVA uses 293.5 seconds to perform one inference on a CIFAR-10 or CIFAR-100 image. The mobile neural network generated by EVA, SqueezeNet, enlarges the layer number of AlexNet, and increases the rescaling operation number to 17 from 14. Each HE operation in SqueezeNet is much slower than that of AlexNet. However, SqueezeNet still reduces the inference latency to 72.7 seconds from 293.5 seconds. This is because SqueezeNet has much less HE operations than AlexNet, leading to a significant performance improvement. However, SqueezeNet generated by EVA is not an optimized mobile neural network architecture.

We use Algorithm~\ref{alg:replace} to find a HE-friendly network based on SqueezeNet. Ours-F$4$ is a network architecture where we use a convolutional layer with 256 input channels, $3\times3$ weight kernels, and 256 output channels, to replace the $4_{th}$ fire module with 256 input channels, $32$ squeezed channels, 128 expanded $1\times 1$ channels, and 128 expanded $3\times 3$ channels. Ours-F$4$ reduces the 72.7-second inference latency of SqueezeNet to 43.8 seconds. Compared to SqueezeNet, Ours-F$4$ is a more RNS-CKKS-friendly network architecture. We further build Ours-F$34$ by replacing both the third and fourth fire modules with two convolutional layers.  Ours-F$34$ achieves shorter inference latency yet even higher inference accuracy. At last, we generate Ours-F$234$ by replacing more fire modules with convolutional layers. But Ours-F$234$ has longer inference latency than Ours-F$34$, which means that the second fire module should not be replaced with a convolution layer. Ours-F$34$ is the best HE-friendly network architecture generated by Algorithm~\ref{alg:replace}.

Coefficient Merging further reduces the number of rescaling operations of Ours-F$34$, thereby decreasing its inference latency. As the Table~\ref{t:SqueezeNet} shows, Ours-F$34$-Merge reduces two rescaling operations in Ours-F$34$ and introduces $20\%$ extra latency reduction. Moreover, Coefficient Merging has no impact on inference accuracy. Overall, compared to the SOTA SqueezeNet built by EVA, Ours-F$34$-Merge reduces the inference latency by $59.3\%$ and improves the inference accuracy by $0.4\%$.

\subsection{InceptionNet}

InceptionNet is a mobile neural network architecture that has more layers than SqueezeNet, so it can achieve higher inference accuracy on both CIFAR-10 and CIFAR-100 datasets. As Table~\ref{t:InceptionNet} shows, 34-layer InceptionNet generated by EVA obtains $83.2\%$ inference accuracy. To guarantee the 128-bit security level, it requires polynomial degree $N=131072$, coefficients modules $Q=2340$, and 39 rescaling operations. And a single PPNN inference of EVA-InceptionNet takes 213.2 seconds.   
	
InceptionNet is not an optimized mobile neural network architecture to achieve such high inference accuracy. Algorithm~\ref{alg:replace} can find more HE-friendly networks based on InceptionNet. As Table~\ref{t:InceptionNet} shows, Ours-I$9$ is one of the architecture generated by Algorithm~\ref{alg:replace}. It replaces the $9_{th}$ inception module by a convolution layer followed by a batch normalization layer. Ours-I$9$ reduces the 213.2-second inference latency of EVA-InceptionNet to only 173.5 seconds. Ours-I$789$ further reduces the inference latency by replacing more inception modules. However, replacing more inception module does not necessarily result in latency reduction. For instance, Ours-I$6789$ suffers from long inference latency than Ours-I$789$. Therefore, Ours-I$789$ is the best HE-friendly mobile neural network architecture found by our Algorithm~\ref{alg:replace}.

We apply Coefficient Merging to further decrease the number of rescaling operations of Ours-I$789$. As Table~\ref{t:InceptionNet} shows, Ours-I$789$-Merge reduces four rescaling operations over Ours-I$789$, and thus introduces $37\%$ latency reduction. Meanwhile, Coefficient Merging does not hurt the inference accuracy. By our search algorithm and Coefficient Merging, Ours-I$789$-Merge reduces the inference latency by $61.2\%$, and improves the inference accuracy by $0.5\%$ over the SOTA InceptionNet compiled by EVA.

\section{Conclusion}

SOTA PPNNs adopt mobile neural network architectures to shorten inference latency and to maintain competitive inference accuracy. We identify that na\"ively applying a mobile neural network architecture on a PPNN increases the multiplicative depth of the entire PPNN, and thus does not necessarily reduce the inference latency. In this paper, we propose HEMET including a simple, greedy HE-friendly network architecture search algorithm and Coefficient Merging to reduce the number of HE operations in a PPNN without increasing the multiplicative depth of the PPNN. Our experimental results show that, compared to SOTA PPNNs, the PPNNs generated by HEMET reduce the inference latency by $59.3\%\sim 61.2\%$, and improves the inference accuracy by $0.4 \sim 0.5\%$.

\section*{Acknowledgements}
The authors would like to thank the anonymous reviewers for their valuable comments and helpful suggestions. This work was partially supported by the National Science Foundation (NSF) through awards CCF-1908992 and CCF-1909509.

\bibliography{he}
\bibliographystyle{icml2021}
	
\end{document}